\documentstyle[11pt,newpasp,twoside,epsf]{article}
\def\edcomment#1{\iffalse\marginpar{\raggedright\sl#1\/}\else\relax\fi}
\reversemarginpar

\begin{document}
\title{Upsilon Andromedae: A Rosetta Stone in Planetary Dynamics}
  \author{E. I. Chiang}
\affil{University of California at Berkeley, Berkeley CA 94720}

\begin{abstract}
We review the orbital dynamics exhibited by
the first extra-solar planetary system discovered, Upsilon
Andromedae. This system is unique in combining all of the
surprising architectural features displayed individually
by extrasolar planetary systems found today: (1) a hot
Jupiter, (2) two planets on highly eccentric orbits, and
(3) a stellar companion. We discuss the system's stability properties
and its possible origin.
Planet-disk interactions seem critical to the emerging story.
\end{abstract}

\section{Introduction}
Upsilon Andromedae ($\upsilon$ And) is a Sun-like star harboring at least three
planetary companions (Butler et al.~1999).
Definitions and current fitted values of planetary orbital parameters are
listed in
Table 1, as supplied by D. Fischer and G. Marcy (2002, personal communication).

\begin{table}
\caption{Fitted Orbital Parameters of Upsilon Andromedae\tablenotemark{a}
\tablenotemark{b}}
\begin{tabular}{cccccccc}
\tableline
Planet & $P$ (days) & $T_{\rm peri}$ (JD) &
$e$	& $\omega$ (deg) & $m \sin i$ ($M_{J}$) & $a$ (AU) \\
\tableline
b & 4.6170 & 12088.710 & 0.0167 & 47.864 & 0.64 & 0.057 \\

c & 241.2473 & 10157.675 & 0.2525 & 245.826 & 1.83 & 0.805 \\

d & 1298.5614 & 13947.615 & 0.3080 & 258.114 & 3.79 & 2.475 \\
\tableline
\tableline
\end{tabular}
\tablenotetext{a}{Based on fitting $N = 207$ radial velocity points.}
\tablenotetext{b}{The orbital period is $P$, the date of pericenter
passage in Julian days is $T_{\rm peri}$, the eccentricity is $e$, the
planetary mass
is $m$ measured in Jupiter masses $M_J$, the inclination of a given
planet's orbit to the plane of the sky is $i$, and the semi-major axis is $a$.
The argument of pericenter, $\omega$, is referred to the sky plane and is
discussed further in the text.}
\end{table}

The quantity $\omega$ in Table 1 is a given
planet's argument of pericenter referred to the plane
of the sky. For an illustration and more details,
see Chiang, Tabachnik, \& Tremaine (2001, hereafter CTT).
Curiously, the $\omega$'s of the outer two planets, c and d,
are presently nearly identical: $\Delta \omega = \omega_d - \omega_c = 12\deg
\pm 5\deg \,
(1\sigma)$. Let us define
$\Theta$ to be the mutual inclination between the two orbit planes
of planets c and d.
If we assume for the moment that $\Theta = 0\deg$, then
the observation that $|\Delta \omega| \ll 1 \,\rm{rad}$ implies
the near perfect alignment of orbital pericenters.

In this review, we address four questions: (1) Is
the near equality of $\omega_d$ and $\omega_c$ coincidental,
or does a dynamical mechanism exist to lock the apsidal lines
together? (2) If a locking mechanism is present, how did
it come to be? (3) What is the origin of the large eccentricities
of planets c and d? (4) What effect does the stellar companion to
$\upsilon$ And (Lowrance, Kirkpatrick, \& Beichman 2002)
have on the dynamics of the planetary system?
Questions (3) and (4) are addressed simultaneously in \S 3.

\section{Existence of Secular Resonance}
The answer to question (1) is that if $\Theta \la 20\deg$,
then the two planets are trapped in an apsidal resonance for
which $|\Delta \omega| \la 38\deg$ (CTT; Chiang \& Murray 2002).
If, however, $\Theta \ga 20\deg$, then the pericenters are
unlocked and today's observation of the smallness of $\Delta \omega$
is accidental.

Figure 1 summaries the results of 855 numerical orbit integrations,
each of duration $10^6 \,\rm{yr}$,
of $\upsilon$ And c and d. The innermost planet, b, is neglected
in these integrations;
its time-averaged potential acts mainly as a small,
static quadrupole moment. Each integration begins with a different
set of five orientation angles consistent with the Doppler velocity data:
$\Theta_0, \Omega_0, \chi_0$---the initial inclination, initial longitude
of ascending node, and initial argument of pericenter, respectively,
of the orbit of planet d referred to
that of planet c---and $\phi, \delta$---the
location of the observer on the celestial sphere centered on $\upsilon$ And.
Simulated systems that begin with $\Theta_0 \la 20\deg$ tend not only
to be stable
but also spend the most time with their $\omega$'s nearly equal.
Such systems inhabit a so-called secular resonance for which
$\Delta \omega$ librates about $0\deg$ with an amplitude
of $\sim$$38\deg$. This behavior can be described analytically
by the Laplace-Lagrange equations (CTT).
Those systems that are stable have $\sin i_c, \sin i_d \ga 0.5$ (data
not shown). Systems at $\Theta_0 \ga 20\deg$ are characterized
by circulating $\Delta \omega$. Those for which
$40\deg \la \Theta_0 \la 140\deg$ are rendered unstable by the
Kozai resonance, by which $e_c$ is driven by
planet d to near unity (CTT).

\begin{figure}
\plotfiddle{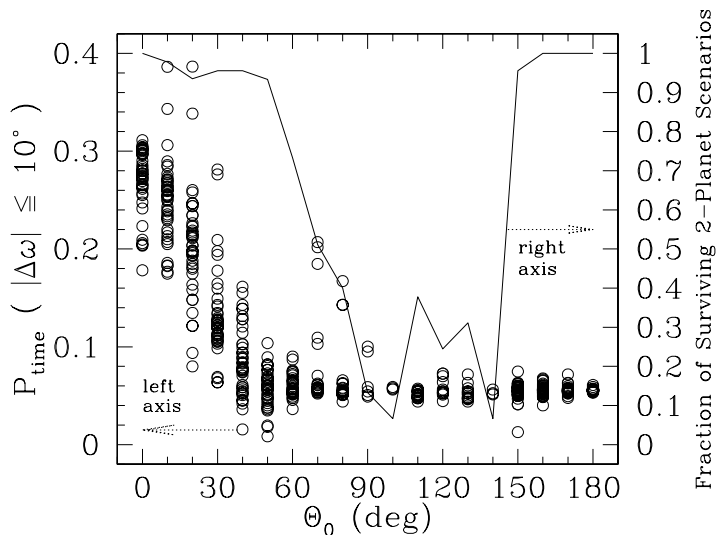}{2.5in}{-90}{35}{35}{-150}{190}
\vspace{0.1in}
\caption{Alignment probability and stability (CTT).
{\it Open circles, left-hand ordinate}: Fraction
of time that surviving, sampled, 2-planet scenarios
spend with $|\Delta \omega| \leq 10\deg$.
{\it Solid line, right-hand ordinate}: Fraction of 2-planet
scenarios that survive the
$10^6 \,\rm{yr}$ duration of the integration.
}
\end{figure}

We submit Figure 1 as our best plausibility argument
that the outer two planets of $\upsilon$ And occupy nearly
co-planar orbits that are
seen nearly edge-on
and that will remain nearly apse-aligned.
This conclusion, in turn, would argue that
these planets formed from a flattened circumstellar disk.
Verification of the smallness of $\Theta$ will probably
have to wait for astrometric measurements of the proper
motion of $\upsilon$ And. These measurements are
currently in progress with the Hubble Space Telescope.

\section{Eccentricity Excitation and Secular Resonance Capture}
That $m_d$ likely exceeds $m_c$ while
$e_d > e_c$ stands at odds with the
idea that gravitational interactions
between planets D and C excited both
$e_d$ and $e_c$ to their present-day values.
We are led to the conclusion that an external
agent---another planet, a star, or
the circumstellar disk from which the
planets formed---directly excited
$e_d$. We favor the third candidate.

\begin{figure}
\plotfiddle{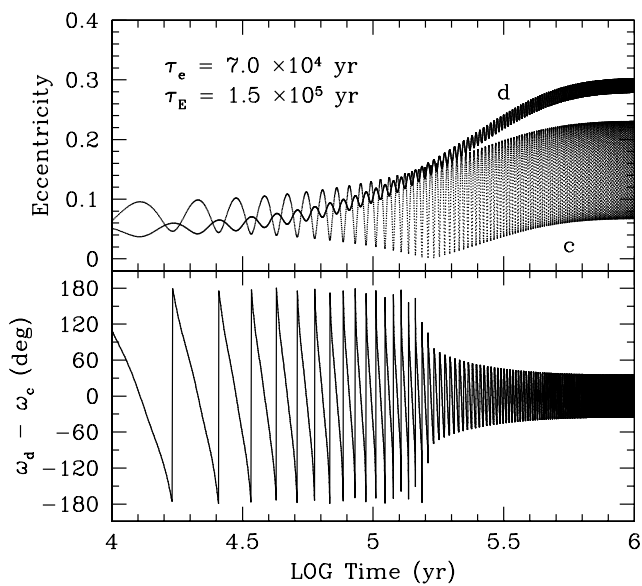}{2.5in}{0}{45}{45}{-150}{-90}
\vspace{0.1in}
\caption{Externally driving the eccentricity of planet d amplifies
the eccentricity of planet c and locks the system into
apsidal resonance. At the end of the integration, the
eccentricities and apsidal longitude difference match
those of $\upsilon$ And today (Chiang \& Murray 2002).
}
\end{figure}

Goldreich \& Sari (2002) describe a finite-amplitude instability
by which a planet's eccentricity can be resonantly excited
by the disk. The timescale for eccentricity amplification
is $e/\dot{e} \sim 7\times 10^4 (10^{-4}/\alpha)^{4/3} (40 M_J/M_D) \,
\rm{yr}$, where $\alpha$ is the usual dimensionless viscosity of the
disk, and $M_D$ is the mass of disk material occupying first-order
Lindblad resonances established by the planet. Suppose
a circumstellar disk sits just exterior to planet d
and excites $e_d$. Provided $\tau_e \equiv e_d/\dot{e_d}$ exceeds
apsidal precession timescales ($\sim$$8000 \, \rm{yr}$),
the (nearly co-planar) system morphs adiabatically through a series of
classical Laplace-Lagrange solutions. Chiang \& Murray (2002)
calculate that the resulting probability of apsidal resonance capture
is 100\%, that $e_c$ grows
by siphoning off $e_d$, and that continued eccentricity
driving damps the apsidal libration amplitude towards zero.
Figure 2 illustrates one sample evolution, in which
$e_d$ is grown from 0 to its present-day value of $\sim$0.3
over a timescale of $\tau_e \sim 10^5 \, \rm{yr}$.

Note that slow, adiabatic growth of $e_d$ is not guaranteed.
The viscosity profile of protoplanetary disks
is poorly understood. If we adopt $\alpha = 10^{-2}$ instead,
then the evolution is impulsive. Malhotra (2002)
computes the evolution in the impulsive limit and calculates
a probability of capture into apsidal resonance of approximately
50\%. Impulsive driving may also result from violent planet-planet
scattering, though the details have yet to be elucidated.

\section{Ups And B: Nemesis or Benign Companion?}
Sitting at a projected separation of 750 AU from $\upsilon$ And
A is $\upsilon$ And B, a proper motion companion of mass
$m_B \sim 0.2 M_{\odot}$ (Lowrance et al.~2002).
Could this companion star be responsible
for exciting the observed eccentricities of planets c and d?
The answer is almost certainly no. The usual Kozai mechanism
for pumping planetary eccentricities probably cannot operate
because apsidal precession rates of the planets
due to planet-planet interactions are likely $\sim$$10^4$ times
faster than those induced by the star (Chiang \& Murray 2002).
The star can never ``get a handle'' on the apses of the planets.

Close encounters between $\upsilon$ And B and planet d are
not important unless the eccentricity of the former's orbit
exceeds 0.98, an unlikely possibility that would threaten
the stability of the entire system.

The impotence of Kozai-type forcing by the star also implies
that it is not the route by which the innermost planet, b,
attained its current close orbit. We cannot, however, rule
out Kozai-type forcing of $e_b$ by planets c and d, though
this would require that the orbit plane of b be once
inclined with respect to those of c and d by more than $\sim$40$\deg$.
``Type II'' migration of planet b induced by a viscous
protoplanetary disk seems a more natural hypothesis (Ward 1997).

\end{document}